\documentclass[reprint,aps,prb]{revtex4-2} 
\usepackage{graphicx}
\usepackage{amsmath}
\begin{document}
\title{Enigmatic factor of 4/3 in electromagnetic momentum of a moving spherical capacitor} 
\author{Ashok K. Singal}
\email{ashokkumar.singal@gmail.com}
\affiliation{Astronomy and Astrophysics Division, Physical Research Laboratory,
Navrangpura, Ahmedabad - 380 009, India }
\date{\today}
\begin{abstract}
The electromagnetic energy-momentum of a moving charged spherical capacitor may be calculated by a 4-vector Lorentz transformation from the energy in the rest frame. However, energy-momentum of the moving system computed directly from electromagnetic fields yields extra terms; in particular a factor of 4/3 in momentum appears, similar to that encountered in the classical electron model, where this enigmatic factor has been a source of scourge for more than a century. There have been many attempts to eliminate this `unwanted' factor, noteworthy among them is a modification in electromagnetic field energy-momentum definition that has entered even standard textbooks. Here it is shown that in a moving charged spherical capacitor, some additional contributions to the electromagnetic energy-momentum arise naturally from electromagnetic forces in system or equivalently from terms in the Maxwell stress tensor; contributions that do not otherwise show up in 4-vector transformations. Furthermore, a similar factor of 4/3 in the momentum of a perfect fluid comprising a randomly moving ultra-relativistic gas molecules or an isotropic photon gas, filling an {\em uncharged} spherical capacitor in motion, appears owing to the contribution of pressure. Thus, genesis of the ``enigmatic'' factor of 4/3 can be traced to pressure or stress whose presence in the system may be of non-electromagnetic origin and where the proposed modifications in energy-momentum definition do not even come into picture. No modifications in the definitions of energy-momentum of moving fluids have ever been required; physics should be the same in electromagnetic case as well, implying there is nothing amiss in the standard definition of electromagnetic energy-momentum.
\end{abstract}
%
\maketitle
\section{Introduction}\label{S1}
There is apparently an extraneous factor of 4/3 in the electromagnetic mass ($m_{\rm el}=4m_0/3$) of a moving classical electron of charge $e$, assumed to be a uniformly charged spherical shell of a  small radius $\epsilon$, with $m_0 =e^2/2\epsilon c^2$ as mass equivalent of energy in electric fields in the rest frame. 
This enigmatic factor of 4/3 first appeared in the works of Abraham \cite{abr05} and Lorentz  \cite{16}, brief historical reviews of which can be found in more recent literature \cite{51,20}. A detailed mathematical analysis of the self-force of a relativistically moving electron from the time-retarded mutual interaction between its constituents was done by Schott in a rather tedious manner~\cite{12}. However, the calculations could be performed much more conveniently in the instantaneous rest frame, where the leading term in self-force, in the approximation of small radius of the electron ($\epsilon \rightarrow 0$), turns out to be $4m_0\dot{v}/3$, with $\dot{v}$ as the non-relativistic acceleration \cite{1,2,3}. 

One arrives at this factor of 4/3 in another way. We can compute the electromagnetic momentum of a moving charge from the electric field energy in rest frame by a Lorenz transformation, treating electromagnetic energy-momentum as a 4-vector. Alternatively, one could directly calculate the momentum in the electromagnetic fields of the moving charge. The latter computation, however, yields an extra factor of 4/3 in the electromagnetic momentum, even for non-relativistic velocities \cite{29}. 

The factor of 4/3 in the electric mass has been a source of confusion for more than a century \cite{Pe82,4,Gr12} and there have been continuous efforts to get rid of it. 
Poincar\'{e} proposed a negative pressure of non-electromagnetic (mechanical!) origin, also known as Poincar\'{e} stresses, inside a classical electron to keep its constituents bound which will otherwise fly apart due to electromagnetic self-repulsion forces in the system \cite{34}. Although it could eliminate the troublesome factor of 4/3, yet it did not explain the exact physics behind its occurrence, how and wherefrom it first arose in the classical electron model. 

It was thought that one should be able to do away with this apparently an extraneous factor, purely within classical electromagnetism itself.
With this in mind, modifications were proposed in the standard definition of the electromagnetic field energy-momentum, purportedly to make them relativistically covariant even when the charges may be present \cite{Ro60,4,Ta63,5}; modifications that have made appearance even in standard text-books of electromagnetism \cite{1,2}. 
For instance, for a charge moving with a velocity ${\bf v}=\mbox{\boldmath $\beta$} c$ with a corresponding Lorentz factor $\gamma=(1-\beta^2)^{-1/2}$ and having electromagnetic fields $\bf E$ and $\bf B$, the energy and momentum in the modified definition are given by the volume integrals \cite{1,5,Ro70,Ko81}
\begin{eqnarray}
\label{eq:84b.00}
{U}&=&\gamma^2\int \frac{E^{2}-B^{2}}{8\pi }\,{\rm d}V\,,\\
\label{eq:84b.00a}
{\bf P}&=&\frac{\gamma^2\mbox{\boldmath $\beta$}}{c}\int \frac{E^{2}-B^{2}}{8\pi }\,{\rm d}V\,,
\end{eqnarray}
where ${U},{\bf P}$, defined above, Lorentz transform as components of a 4-vector. (Note the negative sign in Eqs.~(\ref{eq:84b.00}) and ~(\ref{eq:84b.00a})). A number of arguments have appeared in the literature both for and against the modified definition \cite{31.1,31.2,31.3,31.4,31.5,31.6,Mo95,31}.
In fact, an explanation \cite{5} of the famous Trouton-Noble experiment \cite{59}, based on  
the modified definition (Eq.~(\ref{eq:84b.00})) of the energy in electromagnetic fields has been  cited in support of the modified definition \cite{1}.
However, the real explanation of the Trouton-Noble experiment actually lies in the fact that if a system is in equilibrium in the rest-frame  under the electromagnetic forces and the stabilization forces, it would remain in equilibrium in other frames of reference as well, since both types of forces will transform relativistically in exactly the same manner from one frame to another \cite{58}.

There have been some other suggestions to eliminate the factor of 4/3. One   
suggestion is to implement a transformation of the Poynting theorem to a form, where the terms of self-interaction between charges get eliminated, giving an expression for field momentum compatible with the Lorentz transformation as a 4-vector for total energy-momentum \cite{Kh16}. Mathematically, the essence is not different from the case where Poincar\'{e} stresses cancel the contribution of electromagnetic self-interaction between charges. Another hypothesis that has been put forward is that the problem of 4/3 has a solution if the fields are instantaneous \cite{Ch19}. Again this may not be surprising since the net force of self-interaction arises after all from the {\em time-retarded} mutual interaction between various parts of the charge \cite{1,2,3}. Since it is well-known that the electromagnetic interactions propagate with a finite velocity $c$. So just to solve one particular problem, one could not abandon the idea of a finite speed of propagation in electromagnetism. 
We shall, in fact, show here that well within the standard theory of electromagnetism, without invoking any exotic new  hypotheses, one can explain the factor of 4/3 when contributions to the electromagnetic energy-momentum of all forces within the system are properly taken into account.

It will be shown in section \ref{S3} that a factor of 4/3 appears even in the field momentum of a {\em macroscopic} system like a charged spherical capacitor in motion, and that it arises naturally in a moving system from the contributions of electromagnetic forces between charges, or more formally, from stress terms in the Maxwell stress tensor (section \ref{S4}), terms which do not get represented in the energy-momentum calculated from a 4-vector transformation. Thus this factor of 4/3 has nothing to do with the {\em microscopic} ($\epsilon \rightarrow 0$) structure of a classical electron model, and appears in some form or other in any moving charged system, described by classical electromagnetism and that there is no need, whatsoever, to modify the standard definitions of the electromagnetic field energy-momentum. The question actually is not about the description of some `real' elementary particle within classical electromagnetism, rather the question here is about the mathematical self-consistency of the classical theory itself.

Furthermore, even in a system like a perfect fluid having a bulk motion, pressure makes a similar contribution to the momentum of the system \cite{84}. For instance, in the case of perfect fluid comprising a gas having an ultra-relativistic random motion of molecules or an isotropic photon gas (a.k.a. radiation), where pressure $p$ is related to energy density $u'$ in the rest frame of the fluid as $p= u'/3$, the momentum density ${\bf g}$ of the fluid moving with a non-relativistic {\em bulk} velocity $\bf v$ is ${\bf g}=(u'+p){\bf v}/c^2=4u'{\bf v}/3c^2$. One thus arrives at a factor of 4/3 in the momentum of a spherical capacitor filled with an ultra-relativistic gas or  radiation (section \ref{S6}), and having no electric charges and no consequential electrostatic fields in the rest frame. Thus the ``enigmatic''  factor of 4/3 arises from the contribution of forces or stress whose genesis in the system may be of non-electromagnetic origin, and has less to do with the presence of electric charges or currents within the system. Thus contrary to the conventional wisdom, the 4/3  factor is not something arising exclusively from electric charges in the system and there is nothing amiss in the standard definition of electromagnetic energy-momentum. No modifications in the definitions of energy-momentum of moving fluids have ever been found necessary to be proposed; physics after all should be the same in the case of electromagnetic energy-momentum as well.  
\section{Energy-momentum in the electromagnetic fields of a moving charged spherical capacitor}\label{S2}
We assume the capacitor to comprise two concentric spherical shells, $S_{\rm a}$ and $S_{\rm b}$ of radii $a$ and $b$, with charge $Q$ and $-Q$ distributed uniformly over the surfaces of  $S_{\rm a}$ and $S_{\rm b}$ respectively (Fig. 1). The spherical capacitor has a capacity (in cgs units) \cite{PU85}
\begin{eqnarray}
\label{eq:84b.01}
C= \frac{1}{\frac{1}{a} - \frac{1}{b}}=\frac{ab}{b-a}\:.
\end{eqnarray}
\subsection{Energy-momentum computed from a 4-vector Lorentz transformation from the rest frame}\label{S2A}

The electromagnetic energy and momentum of the capacitor system are computed from the volume integrals of the energy and momentum densities in the electromagnetic fields
\begin{eqnarray}
\label{eq:84b.02}
{U}_{\rm f}&=&\int \frac{E^{2}+B^{2}}{8\pi }\,{\rm d}V\,,\nonumber\\
{\bf P}_{\rm f}&=&\int \frac{{\bf E} \times {\bf B}}{4\pi c}\,{\rm d}V\,.
\end{eqnarray}

Let the capacitor be stationary in an inertial frame ${\cal K}'$, called the rest frame. The electric field is zero inside the inner sphere $S_{\rm a}$ as well as  outside the outer sphere $S_{\rm b}$, and there will only be a radial electric field at $r'$ ($a<r'<b$) in rest frame $\cal K'$. From the position vector ${\bf r}'=x' \hat{\bf x}+y' \hat{\bf y}+z' \hat{\bf z}$, we can express ${\bf E}'= Q{\bf r}'/r'^3=Q (x' \hat{\bf x}+y' \hat{\bf y}+z' \hat{\bf z})/{r'^{3}}$ for ($a<r'<b$) and zero elsewhere. The magnetic field of course is zero everywhere in ${\cal K}'$. 

The volume integral of the energy density in rest frame ${\cal K}'$ is easily calculated to be 
\begin{eqnarray}
\label{eq:84b.03}
{U}'&=&\int_{a} ^{b} \frac{E'^{2}}{8\pi }\,{\rm d}V' =\int_{a} ^{b} \frac{Q^{2}}{8\pi r'^4 }\,4\pi r'^2 {\rm d}r'\nonumber\\
&=& \frac{Q^2}{2}\left(\frac{1}{a} - \frac{1}{b}\right)= \frac{Q^2}{2C}\:.
\end{eqnarray}
The electromagnetic momentum $\bf P'$ is zero in rest frame ${\cal K}'$ as the magnetic field is zero throughout in that frame. If the system has an acceleration $c{\rm d}\mbox{\boldmath $\beta$'}/{\rm d}t'$ in the instantaneous rest frame, then we can write ${\rm d}{\bf P}'/{\rm d}t'=(U'/c){\rm d}\mbox{\boldmath $\beta$'}/{\rm d}t'$. The rate of change of energy ${U}'$ in the rest frame  is nil, i.e., ${\rm d}U'/{\rm d}t'=0$.
\begin{figure}
\begin{center}
\includegraphics[width=7cm]{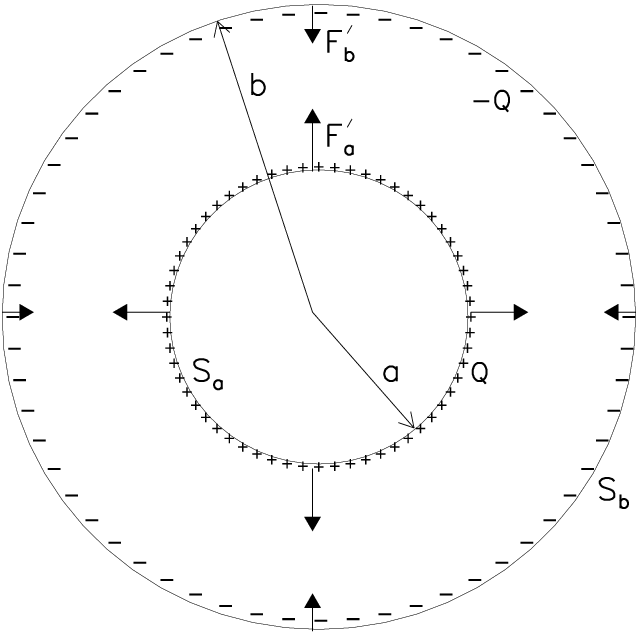}
\caption{A schematic diagram of a charged spherical capacitor in its rest frame ${\cal K}'$. The capacitor comprises an inner sphere $S_{\rm a}$ of radius $a$ and an outer sphere $S_{\rm b}$ of radius $b$, with charges $Q$ and $-Q$ distributed uniformly on the inner and outer surfaces respectively. The electric field is radial, ${\bf E}'= Q{\bf r'}/r'^3$, for $a<r'<b$ but is zero both within $S_{\rm a}$ and without $S_{\rm b}$. ${F}'_{\rm a}$ represents a radially outward electric force per unit area on surface $S_{\rm a}$, while the electric force per unit area on surface $S_{\rm b}$ is ${F}'_{\rm b}$, which, because of the negative charge density on $S_{\rm b}$, is in a radially inward direction.}
\label{F1}
\end{center}
\end{figure}

Let in the lab frame  ${\cal K}$, the rest frame ${\cal K}'$ of the capacitor, is moving say, along $x$-axis, with velocity ${\bf v}=\mbox{\boldmath $\beta$} c$ and a Lorentz factor $\gamma=(1-\beta^2)^{-1/2}$. Now, if energy and momentum are Lorentz transformed as components of a 4-vector from the rest frame ${\cal K}'$ to the lab frame $\cal K$, we get 
\begin{eqnarray}
\label{eq:84b.38.1}
{U}&=&\gamma  {U}'\:.\\
\label{eq:84b.39.1}
{\bf P}&=&{U}' \frac{\gamma \mbox{\boldmath $\beta$}}{c}\:.
\end{eqnarray}
In frame $\cal K$, energy and momentum are, accordingly, related by
\begin{eqnarray}
\label{eq:84b.39}
{\bf P}&=&{U} \frac{\mbox{\boldmath $\beta$}}{c}\:.
\end{eqnarray}
Because of an acceleration $c{\rm d}\mbox{\boldmath $\beta$}/{\rm d}t$  of the system in the lab frame, 
we have 
\begin{eqnarray}
\label{eq:84b.38.0}
\frac{{{\rm d}{\gamma}}}{{\rm d}t}&=&\gamma^3\mbox{\boldmath $\beta$}\cdot\frac{{\rm d}\mbox{\boldmath $\beta$}}{{\rm d}t}\,,\\
\label{eq:84b.38.00}
\frac{{{\rm d}({\gamma}\mbox{\boldmath $\beta$}})}{{\rm d}t}&=&\gamma^3\Big(\mbox{\boldmath $\beta$}\cdot \frac{{\rm d}\mbox{\boldmath $\beta$}}{{\rm d}t}\Big)\mbox{\boldmath $\beta$}
+\gamma \frac{{\rm d}\mbox{\boldmath $\beta$}}{{\rm d}t}\,. 
\end{eqnarray}
From  Eq.~(\ref{eq:84b.38.1}) and  Eq.~(\ref{eq:84b.39.1}), with the help of Eqs.~(\ref{eq:84b.38.0}) and  (\ref{eq:84b.38.00}), we can write
\begin{eqnarray}
\label{eq:84b.38.2}
\frac{{\rm d}{U}}{{\rm d}t}&=&U'\frac{{\rm d}{\gamma}}{{\rm d}t}
=U' \gamma^3\mbox{\boldmath $\beta$}\cdot \frac{{\rm d}\mbox{\boldmath $\beta$}}{{\rm d}t},\\
\label{eq:84b.38.3}
\frac{{\rm d}{\bf P}}{{\rm d}t}&=&U'\frac{{\rm d}({\gamma}\mbox{\boldmath $\beta$})}{c\:{\rm d}t}=U'\gamma^3\Big(\mbox{\boldmath $\beta$}\cdot \frac{{\rm d}\mbox{\boldmath $\beta$}}{c\:{\rm d}t}\Big)\mbox{\boldmath $\beta$}
+U' \gamma \frac{{\rm d}\mbox{\boldmath $\beta$}}{c\:{\rm d}t}. 
\end{eqnarray}
Then from  Eq.~(\ref{eq:84b.38.2}) and  Eq.~(\ref{eq:84b.38.3}), using $\gamma^2\beta^2+1=\gamma^2$, we get
\begin{eqnarray}
\label{eq:84b.38}
\frac{{\rm d}{U}}{{\rm d}t}=\frac{{\rm d}{\bf P}}{{\rm d}t} \cdot {\mbox{\boldmath $\beta$}}{c}\:.
\end{eqnarray}
This can be understood in another way. Because of the force $\bf F={{\rm d}{\bf P}}/{{\rm d}t}$ in the lab frame ${\cal K}$, where the system is moving with a velocity ${\bf v}=\mbox{\boldmath $\beta$}{c}$, energy of the system is changing at a rate ${\bf F\cdot v}$. 
\subsection{Energy-momentum computed directly from electromagnetic fields of a moving charged capacitor}\label{S2B}

One can also calculate the energy and momentum of the system in the lab frame $\cal K$ directly from the fields. 
The parallel component of electric field does not change, i.e. $E_{\|}= E'_{\|}$. But the perpendicular component is higher in the lab frame $\cal K$ by the Lorentz factor $\gamma$, i.e. $E_{\bot}=\gamma E'_{\bot}$.  Magnetic field is given by ${\bf B}=\mbox{\boldmath $\beta$}\times {\bf E}$, i.e., $B=\beta E_{\bot}$ \cite{PU85}. 

The volume integrals of the energy and momentum densities of the electromagnetic fields of the  capacitor in frame ${\cal K}$, where it is moving with a velocity $\bf v$, say along $x$-axis, and the volume between the spheres is consequently Lorentz contracted, 
could be computed more conveniently by transposing to the rest frame as 
\begin{eqnarray}
\label{eq:84b.33}
{U}_{\rm f}&=&\int \frac{E^{2}+B^{2}}{8\pi }\,{\rm d}V=\int \frac{E'^{2}_{\|}+(\gamma^{2}+\gamma^{2}\beta^2) E'^{2}_{\bot}}{8\pi\gamma}\,{\rm d}V',\nonumber\\
{\bf P}_{\rm f}&=&\int \frac{{\bf E} \times {\bf B}}{4\pi c}\,{\rm d}V=\int \frac{\gamma^{2}\mbox{\boldmath $\beta$} E'^{2}_{\bot}}{4\pi c \gamma}\,{\rm d}V'.
\end{eqnarray}
where, we have substituted ${\rm d}V={\rm d}V'/\gamma$, to account for the Lorentz contraction.

Now because of the circular-cylindrical symmetry of the system about the x-axis, the direction of motion, we have
\begin{eqnarray}
\label{eq:84b.34}
 \int E'^{2}_{\|}dV'=\frac{1}{2}\int E'^{2}_{\bot}dV'=\frac{1}{3}\int E'^{2}\,{\rm d}V'.
\end{eqnarray}
Then, we get for the field energy and momentum in ${\cal K}$
\begin{eqnarray}
\label{eq:84b.35}
{U}_{\rm f}&=&\gamma \left(1+\frac{\beta^2}{3}\right)\frac{Q^2}{2}\left(\frac{1}{a} - \frac{1}{b}\right)\nonumber\\
&=&\gamma \left(1+\frac{\beta^2}{3}\right)\frac{Q^2}{2C}=\frac{4}{3}\,{U}'\gamma-\frac{{U}'}{3\gamma}\:,\\
\label{eq:84b.37}
{\bf P}_{\rm f}&=&\frac{4}{3}\frac{\gamma \mbox{\boldmath $\beta$}}{c}\frac{Q^2}{2}\left(\frac{1}{a} - \frac{1}{b}\right)=\frac{4}{3}U'\frac{\gamma \mbox{\boldmath $\beta$}}{c}
\:.
\end{eqnarray}

From Eq.~(\ref{eq:84b.35}) and  Eq.~(\ref{eq:84b.37}), with the help of Eqs.~(\ref{eq:84b.38.0}) and  (\ref{eq:84b.38.00}), we can write
\begin{eqnarray}
\label{eq:84b.35.1}
\frac{{\rm d}{U_{\rm f}}}{{\rm d}t}&=&\frac{4}{3}U'\frac{{\rm d}{\gamma}}{{\rm d}t}-\,\frac{{U}'}{3}\frac{{\rm d}}{{\rm d}t}\left(\frac{1}{\gamma}\right)\nonumber\\
&=&\frac{4}{3}U' \gamma^3\mbox{\boldmath $\beta$}\cdot \frac{{\rm d}\mbox{\boldmath $\beta$}}{{\rm d}t}-\,\frac{{U}'}{3}\frac{{\rm d}}{{\rm d}t}\left(\frac{1}{\gamma}\right)\:,\\
\label{eq:84b.37.1}
\frac{{\rm d}{\bf P}_{\rm f}}{{\rm d}t}&=&\frac{4}{3}U'\frac{{\rm d}({\gamma}\mbox{\boldmath $\beta$})}{c\:{\rm d}t}\nonumber\\
&=&\frac{4}{3}U'\gamma^3\Big(\mbox{\boldmath $\beta$}\cdot \frac{{\rm d}\mbox{\boldmath $\beta$}}{c\:{\rm d}t}\Big)\mbox{\boldmath $\beta$}
+\frac{4}{3}U' \gamma \frac{{\rm d}\mbox{\boldmath $\beta$}}{c\:{\rm d}t}\:. 
\end{eqnarray}

Then from  Eq.~(\ref{eq:84b.35.1}) and  Eq.~(\ref{eq:84b.37.1}), using $\gamma^2\beta^2+1=\gamma^2$, we get
\begin{equation}
\label{eq:84b.40a}
\frac{{\rm d}{U}_{\rm f}}{{\rm d}t}=\frac{{\rm d}{\bf P}_{\rm f}}{{\rm d}t}\cdot{\mbox{\boldmath $\beta$}}{c}-\,\frac{{U}'}{3}\frac{{\rm d}}{{\rm d}t}\left(\frac{1}{\gamma}\right)\,.
 \end{equation}

There are a couple of noteworthy points here. Compared to Eq.~(\ref{eq:84b.38}), Eq.~(\ref{eq:84b.40a}) has an additional term, the second term on the right hand side. As we will demonstrate in section~(\ref{S3B}, Eq.~(\ref{eq:84b.3q})), this additional term in Eq.~(\ref{eq:84b.40a}) represents the rate of work being done during Lorentz contraction of the system against the electromagnetic forces in the charged system. Also, using Eq.~(\ref{eq:84b.35}) to write Eq.~(\ref{eq:84b.37}) as 
\begin{eqnarray}
\label{eq:84b.37.2}
{\bf P}_{\rm f}&=&{U}_{\rm f}\frac{\mbox{\boldmath $\beta$}}{c}+\frac{{U}'}{3\gamma}\frac{ \mbox{\boldmath $\beta$}}{c}\,.
\end{eqnarray}
we note that compared to Eq.~(\ref{eq:84b.39}), there is an additional term in field momentum, the second term on the right hand side, in Eq.~(\ref{eq:84b.37.2}).  It is this term, whose origin will be explored in detail in section~(\ref{S3A}, Eq.~(\ref{eq:84b.54a})), that is responsible for the enigmatic factor of 4/3 in electromagnetic momentum of a moving charged system.

Further, for a non-relativistic motion ($v\ll c$) of the spherical capacitor, keeping first order terms in $\beta$ with $\gamma \approx 1$ in Eqs.~(\ref{eq:84b.35}) and (\ref{eq:84b.37}), while we do have ${U}_{\rm f}= {U}'$, implying no change, to that order, in the electromagnetic field energy of the system from the rest frame, however, we still get ${\bf P}_{\rm f}={4}{U}'{\mbox{\boldmath $\beta$}}/3{c}$, consistent with the view that the factor of 4/3, encountered historically in the  electromagnetic mass of a classical electron model, is of a non-relativistic origin \cite{29}. It can be noted that as outer sphere of the charged spherical capacitor becomes very large ($b\rightarrow \infty$), Eqs.~(\ref{eq:84b.03}), (\ref{eq:84b.35}) as well as  (\ref{eq:84b.37}), turn into those of a charged sphere of radius $a$, as in the classical electron model \cite{15,Mo11}.
\section{Energy-momentum of a moving charged spherical capacitor from electromagnetic forces on the spherical surfaces}\label{S3}
Charge $Q$ distributed uniformly over the surface area $4\pi a^{2}$ of the inner sphere $S_{\rm a}$ of radius $a$, creates a surface charge density $\sigma_{\rm a}=Q/4\pi a^{2}$, which gives rise to an outward radial electric field, ${\bf E}'_{\rm a}= 4\pi\sigma_{\rm a}{\bf n}$, on the surface of $S_{\rm a}$ \cite{PU85}, where ${\bf n}$ is an outward unit radial vector on the spherical surface. Electric field inside the sphere is zero. Of course, due to the charge of $S_{\rm b}$ there is no electric field on the surface of  $S_{\rm a}$. 
Accordingly, there is an average \cite{PU85}, radially outward force, ${\bf F}'_{\rm a}= 2\pi\sigma^2_{\rm a}{\bf n}=(Q^2/8\pi a^{4}){\bf n}$, per unit area of the surface of $S_{\rm a}$. 

On the other hand, the electric field on the surface $S_{\rm b}$ is the field $Q/b^{2}\hat{\bf r}$ due to charge $Q$ on $S_{\rm a}$ as well as a radial inward field $-Q/b^{2}\hat{\bf r}$ due to charge density $\sigma_{\rm b}=-Q/4\pi b^{2}$ on outside of surface $S_{\rm b}$, while the field inside surface $S_{\rm b}$ due to charge on $S_{\rm b}$ is zero. This results in a net average radially outward electric field ${\bf E}'_{\rm b}= Q/2 b^{2}{\bf n}$, on $S_{\rm b}$. Due to the  negative charge density on $S_{\rm b}$ 
there is, accordingly, a radial inward force, ${\bf F}'_{\rm b}= -(Q^2/8\pi b^{4}){\bf n}$, on the unit area surface of the outer sphere $S_{\rm b}$.

There are two subtle factors that make additional contributions to the energy and momentum of the system of a moving sphere in the presence of the electromagnetic forces, ${\bf F}'_{\rm a}$ and ${\bf F}'_{\rm b}$.
We shall now demonstrate how these forces on the moving spherical capacitor surfaces in the lab frame $\cal K$, give rise to momentum and energy terms that indeed are the additional terms encountered in Eqs.~(\ref{eq:84b.37}) and (\ref{eq:84b.40a}). 
\subsection{Momentum contribution from work done by electromagnetic forces in moving charged spherical capacitor}\label{S3A}
We first compute the contribution of force ${\bf F}'_{\rm a}$ on the inner sphere $S_{\rm a}$.
For that we consider two infinitesimal surface elements in the shape of circular rings, $R_1$ and $R_2$, each of radius $a \sin\theta$ and of angular width $d\theta $ on two opposite hemispheres, $\Sigma_1$ and $\Sigma_2$ (Fig~2). Each ring, $R_1$ as well as $R_2$, has an infinitesimal surface area ${\rm d}S=2\pi a^{2} \sin\theta {\rm d}\theta$ and the two rings, as seen in frame ${\cal K}'$, are a distance $l'=2a\cos \theta$ apart, along the direction of motion. 
Net electromagnetic force on the ring $R_1$ is along $x$-axis and is of magnitude
\begin{eqnarray}
\label{eq:84b.52}
2\pi \sigma_{\rm a}^{2}\cos \theta\:{\rm d}S=\frac {Q^2}{4a^{2}} \cos \theta\sin\theta {\rm d}\theta\:.
\end{eqnarray}
There is an equal and opposite force on a symmetrically placed ring $R_2$ in the opposite hemisphere $\Sigma_2$ (Fig.~2). The net force on $R_1$ or $R_2$, being parallel to the direction of motion of the rings in frame $\cal K$, remains unaltered during Lorentz transformation from rest frame ${\cal K}'$ to the lab frame $\cal K$. However, the distance between $R_1$ and $R_2$ is Lorentz contracted in the lab frame $\cal K$ to a value $l=l'/\gamma$.

Since in the frame $\cal K$, the ring $R_1$ is moving with a velocity $\bf v$, the charged ring $R_1$ of a thin surface thickness, say $\Delta r$, and thus of a volume charge density $\sigma/\Delta r$ within the ring, carries a current density ${\bf j}= \sigma {\bf v}/\Delta r$. Then the electric field ${\bf E}'_{\rm a}= 2\pi \sigma {\bf n}$ within the ring thickness, which is an average of the field just outside and inside the sphere $S_{\rm a}$ \cite{PU85}, does work at a temporal rate, $ {\bf E}'_{\rm a}\cdot {\bf j}$  per unit volume, or $ {\bf E}'_{\rm a}\cdot {\bf j}\:\Delta r=2\pi \sigma^2 v\cos \theta $ on per unit surface area of the ring. Thus work is being done by the electromagnetic force on the moving ring at a rate $({Q^2}/{4a^{2}}) v \cos \theta\sin\theta {\rm d}\theta$. 

\begin{figure}[t]
\begin{center}
	\includegraphics[width=7cm]{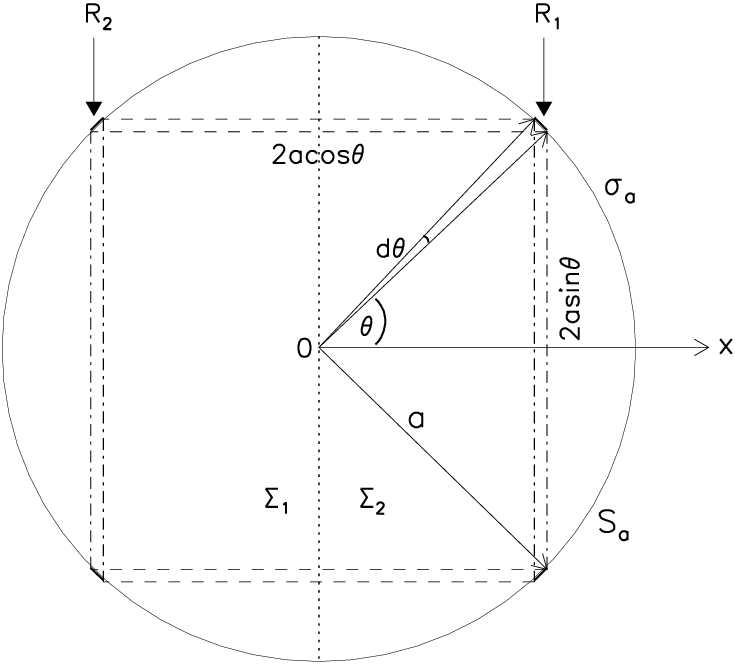}
\caption{The sphere $S_{\rm a}$ of radius $a$ has a uniform surface charge density $\sigma_{\rm a}$. 
The vertical dotted line passing through the center O represents a central plane that divides the sphere $S_{\rm a}$ into two equal hemispheres $\Sigma_1$ and $\Sigma_2$.
Dash-dotted lines represent two symmetrically placed circular rings, $R_1$ and $R_2$, each of radius $a \sin \theta$ and of angular width ${\rm d}\theta$, lying on two opposite sides of the spherical surface, separated by a distance $2a \cos \theta$, represented by horizontal dashed lines along the $x$ direction.}
\label{F2}
\end{center}
\end{figure}

Now at the same rate an opposite amount of work is being done by the electromagnetic force at ring $R_2$. 
Thus the energy increase at this rate due to electromagnetic interaction at $R_1$ is at the cost of an equal decrease at the same rate  at $R_2$, a distance $l=2a\,\cos\theta \:/\gamma$ away in the lab frame $\cal K$. Accordingly, there is an electromagnetic energy flow from $R_2$ to $R_1$ at a rate $({Q^2}/{2a \gamma}) v \cos^2 \theta\sin\theta {\rm d}\theta$
through a cylindrical cross-section between $R_2$ and $R_1$, implying an additional momentum in the system, 

Here a question could justifiably be raised, where is the source (or sink) of this energy since both $R_1$ and  $R_2$ continue to move with constant velocity?  Actually there are other forces, of non-electromagnetic (mechanical!) origin, necessarily present in the system to keep it in equilibrium. For instance, in the case of charged spherical capacitor, there will be binding forces that keep the charges confined (`glued!') to their respective locations on the capacitor plate surfaces in spite of the net electrostatic force on them that would otherwise make the charges fly off the capacitor plate surfaces. 
The Contribution of such non-electromagnetic forces, first proposed by Poincar\'{e} to keep a charged system in equilibrium \cite{34}, to the total momentum of the system will be discussed in section \ref{S5}. However, since these non-electromagnetic interactions {\em do not get included} in the expressions for electromagnetic fields, whether in the rest frame  ${\cal K}'$ or in the lab frame $\cal K$, therefore the electromagnetic field momentum, computed from {\em electromagnetic fields}, is still given by Eq.~(\ref{eq:84b.37}).

Integration over the total cross-section of the sphere yields a net additional momentum due to this energy flow between hemispheres $\Sigma_1$ and $\Sigma_2$ of $S_{\rm a}$. 
\begin{eqnarray}
\label{eq:84b.53a}
\Delta{\bf P}_{\rm a}=\int_0^{\pi/2}\frac {Q^2}{2a\gamma} \frac{\bf v}{c^{2}} \cos^2 \theta\sin\theta {\rm d}\theta=\frac {Q^2}{2a\gamma}\frac{\bf v}{3c^{2}}\:.
\end{eqnarray}
Similarly, due to the outer sphere $S_{\rm b}$, where ${\bf F}'_{\rm b}$ is radially inward, there is an additional momentum 
\begin{eqnarray}
\label{eq:84b.54}
\Delta{\bf P}_{\rm b}=-\frac {Q^2}{2b\gamma}\frac{\bf v}{3c^{2}}\:.
\end{eqnarray}
Together, the additional momentum in the system is
\begin{eqnarray}
\label{eq:84b.54a}
\Delta{\bf P}=\Delta{\bf P}_{\rm a}+\Delta{\bf P}_{\rm b}= \frac {Q^2}{2\gamma}\left(\frac{1}{a} - \frac{1}{b}\right)\frac{v}{3c^{2}}=\frac{{U}'}{3\gamma}\frac{ \mbox{\boldmath $\beta$}}{c}\:,
\end{eqnarray}
which is the additional term in the electromagnetic field momentum (Eq.~(\ref{eq:84b.37.2})).

Thus, the second term on the right hand side in Eq.~(\ref{eq:84b.37.2}) of electromagnetic momentum can be identified with the energy flow between different parts of the system along the direction of motion, with electromagnetic forces doing different work on different parts, thereby explaining the genesis of the enigmatic factor of 4/3 in the electromagnetic momentum in such a moving charged systems.
\subsection{Energy contribution from electromagnetic forces during work done against Lorentz contraction}\label{S3B}
Suppose in the lab frame ${\cal K}$, where the capacitor system is moving with velocity $v=\beta c$ along $x$-axis, the capacitor gets a velocity increment $c\Delta\beta$ due to an acceleration component along the $x$-axis, and a corresponding increment in the Lorentz factor $\Delta\gamma\approx\gamma^3\beta\Delta\beta$, in a small time interval $\Delta t$. Because of the increase in the Lorentz factor, the distance between $R_1$ and $R_2$ decreases in frame ${\cal K}$ by 
\begin{eqnarray}
\label{eq:84b.40c}
\Delta l=\frac{l'}{\gamma}-\frac{l'}{\gamma+\Delta\gamma}\approx\frac{l'\Delta\gamma}{\gamma^2}=l'\gamma\beta\Delta\beta\:,
\end{eqnarray}
where we have assumed that the distance $l'$ between rings  $R_1$ and $R_2$ remains unaltered, maintaining the shape of the capacitor as spherical, in the instantaneous rest frame.

Now, during a change in the Lorentz contraction of the sphere  $S_{\rm a}$, work would be done against  the $x$ component of the outward force per unit area, ${\bf F}'_{\rm a}$, which is equivalent to an outward pressure on the surface  $S_{\rm a}$. Energy will be spent during a contraction of the sphere against this pressure. This energy increment will be in addition to the changing energy of the system even in the absence of the force ${\bf F}'_{\rm a}$,
as in Eq.~(\ref{eq:84b.38}). 

It should be noted that the Lorentz contraction of a moving finite object (which may be a sphere or any other object like a rod) is a real contraction of its length in space, and occurs due to a differential movement between various parts of the object along its direction of motion~\cite{83}.
To comprehend it we consider a rod of length $l'$ in its instantaneous rest frame ${\cal K}'$,  where an equal push is given to all parts of the rod simultaneously, so as to keep its length constant in the instantaneous rest frame. 

Let the rod, as a result, gain a small velocity increment $c\Delta\beta$ along $x$-axis, in lab frame ${\cal K}$. 
Now even if the push given to two ends of the rod is simultaneous in ${\cal K}'$, in frame ${\cal K}$, where the rod is moving with velocity $c\beta$, the trailing end of the rod will get the velocity increment $c\Delta\beta$ {\em before} the leading end by a tiny time interval $l'\gamma\beta/c$, thereby giving rise to a differential motion and a consequential shortening of the distance between the two ends by $l'\gamma\beta\Delta\beta$, the same as obtained in Eq.~(\ref{eq:84b.40c}), due to change in Lorentz contraction of the rod length. Thus we see that a Lorentz contraction is a real contraction in length due to relative movement of the system parts along the  direction of motion. If there are outward forces on the rod ends, energy will be spent against these outward forces during the moving closer of two ends of the rod by a distance $l'\gamma\beta\Delta\beta$, in the lab frame.

Now in the case of rings $R_1$ and $R_2$, during a changing Lorentz contraction, distance between both rings reduces at a rate 
\begin{equation}
\label{eq:84b.55}
-2a\,\cos\theta \,\frac{{\rm d}}{{\rm d}t}\left(\frac{1}{\gamma}\right)\,,
\end{equation}
Using Eq.~(\ref{eq:84b.52}), 
the rate of work done due to Lorentz contraction against forces on rings $R_1$ and $R_2$ is 
\begin{equation}
\label{eq:84b.56}
-\frac {Q^2}{2a}\frac{{\rm d}}{{\rm d}t}\left(\frac
{1}{\gamma}\right)\, \cos^2 \theta\sin\theta {\rm d}\theta\,.
\end{equation}
After an integration over $\theta$ from $0$ to $\pi/2$, we get the rate of  total work
being done on the charged sphere $S_{\rm a}$ due to Lorentz contraction as
\begin{equation}
\label{eq:84b.57}
-\frac {Q^2}{2a}\frac{{\rm d}}{{\rm d}t}\left(\frac
{1}{\gamma}\right)\,\int_0^{\pi/2} \cos^2 \theta\sin\theta {\rm d}\theta=
-\,\frac{Q^2}{6a}\frac{{\rm d}}{{\rm d}t}\left(\frac{1}{\gamma}\right)\,.
\end{equation}

Similarly, we get the rate of total work being done on the outer sphere $S_{\rm b}$ due to Lorentz contraction as 
\begin{eqnarray}
\label{eq:84b.43}
\frac{Q^2}{6b}\frac{{\rm d}}{{\rm d}t}\left(\frac{1}{\gamma}\right)\,,
\end{eqnarray} 
Then the rate of total work being done on charged spherical capacitor due to changing Lorentz contraction is
\begin{eqnarray}
\label{eq:84b.3q}
-\frac{Q^2}{6}\left[\frac{1}{a}-\frac{1}{b}\right]\frac{{\rm d}}{{\rm d}t}\left(\frac{1}{\gamma}\right)=-\frac{{U}'}{3}\frac{{\rm d}}{{\rm d}t}\left(\frac{1}{\gamma}\right)\,,
\end{eqnarray} 
which agree with the additional term in Eq.~(\ref{eq:84b.40a}). Thus the additional term in Eq.~(\ref{eq:84b.40a}) is due to rate of work being during changing Lorentz contraction of the system against the electromagnetic forces in the charged system. 
\section{Energy-momentum of a moving charged spherical capacitor from electromagnetic stress-energy tensor}\label{S4}
One can compute the energy-momentum of a charged system, more formally, from the electromagnetic stress-energy tensor ${T}^{\mu\nu}$ \cite{1,2,MTW73,Sc85}, where one gets the electromagnetic energy density from the electromagnetic stress-energy tensor as
\begin{eqnarray}
\label{eq:84a.24}
{T}^{00}={\frac {1}{8\pi }}(E^{2}+B^{2})\,,
\end{eqnarray}
while the energy flux ($c{T}^{0i}$) in $i$th direction, as well as $\it i$th component of the electromagnetic momentum density (${T}^{i0}/c$) are written as 
\begin{eqnarray}
\label{eq:84a.25}
{T}^{i0} = {T}^{0i} ={\frac {1}{4\pi}}(\mathbf {E} \times \mathbf {B})^i\,. 
\end{eqnarray}
The symmetric Maxwell stress tensor, given by
\begin{eqnarray}
\label{eq:84a.26}
{\displaystyle {T}^{ij}=\frac {1}{8\pi }\left[\left(E^{2}+B^{2}\right)\delta^{ij}-2(E^{i}E^{j}+B^{i}B^{j})\right]}
\end{eqnarray}
yields $\it i$th component of the momentum flux in $\it j$th direction \cite{1,2,MTW73}. Kronecker delta, $\delta^{ij}=1$ for $i=j$ and zero otherwise. We throughout follow the convention where Greek letters $\alpha, \mu, \nu, \xi$ take the values $0,1,2,3$ while Latin letters $i,j$ take the values $1,2,3$. 

The electromagnetic momentum of the charged spherical capacitor in lab frame $\cal K$  can be determined from a Lorentz transformation of the stress-energy tensor from the rest frame $\cal K'$, where ${\bf E}'= Q (x' \hat{\bf x}+y' \hat{\bf y}+z' \hat{\bf z})/{r'^{3}}$ (for $a<r'<b$) and ${\bf B}=0$ everywhere and accordingly the energy flux and the electromagnetic momentum are zero throughout.
Employing Eqs.~(\ref{eq:84a.24}), (\ref{eq:84a.25}) and (\ref{eq:84a.26}), the stress-energy tensor of the charged spherical capacitor in the rest frame $\cal K'$ can then be expressed as 
\begin{eqnarray}
\label{eq:84b.3g}
{\displaystyle T'^{\mu' \nu'}
{\displaystyle =\frac{Q^2}{8\pi r'^6}{\begin{bmatrix}r'^2&0&0&0\\0&r'^2-2x'^2&-2x'y'&-2x'z'\\0&-2y'x'&r'^2-2y'^2&-2y'z'\\0&-2z'x'&-2z'y'&r'^2-2z'^2\end{bmatrix}}}}\:,
\end{eqnarray} 
for $a<r'<b$, elsewhere all terms are zero. 

In order to determine the electromagnetic momentum of a system in the lab frame  ${\cal K}$, we can perform a Lorentz transformation of the stress-energy tensor  
\begin{eqnarray}
\label{eq:84b.1b}
{\displaystyle {T}^{\mu\nu}={\Lambda ^{\mu}}_{\alpha'}{\Lambda ^{\nu}}_{\xi'}T'^{\alpha' \xi'}\,.} 
\end{eqnarray}
where the matrix for Lorentz transformation between  ${\cal K}'$ and  ${\cal K}$ is given by \cite{MTW73}  
\begin{eqnarray}
\label{eq:84b.1a}
{\displaystyle \Lambda^{\mu}_{\alpha'}=
 {\begin{bmatrix}\gamma&\gamma \beta&0&0\\\gamma \beta&\gamma&0&0\\0&0&1&0\\0&0&0&1\end{bmatrix}}}\:.
\end{eqnarray}

From Eq.~(\ref{eq:84b.3g}), applying Eqs.~(\ref{eq:84b.1b}) and (\ref{eq:84b.1a}), after a bit long but straightforward computations, we get the energy and momentum densities in the lab frame as 
\begin{eqnarray}
\label{eq:84b.3k}
T^{00} &=&\gamma^2(T'^{00}+T'^{11}\beta^2)\nonumber\,,\\ 
 &=&\frac{Q^2\gamma^2[1+(1-2x'^2/r'^2)\beta^2]}{8\pi r'^4}\,,\\
\label{eq:84b.3l}
\frac{1}{c}T^{10} &=&\big(T'^{00} + T'^{11}\big)\frac{\gamma^2\beta}{c}\nonumber\,,\\
&=&\frac{Q^2[1+(1-2x'^2/r'^2)]}{8\pi r'^4}\frac{\gamma^2\beta}{c}\,.
\end{eqnarray} 

A volume integral gives electromagnetic energy-momentum of the charged spherical capacitor in $\cal K$ as
\begin{eqnarray}
	{U}_{\rm em}&=&\int T^{00} {{\rm d}V}= \gamma^2\int [T'^{00}+ T'^{11}\beta^2] \frac{{\rm d}V'}{\gamma}
	\nonumber\\
	&=&\frac{Q^2\gamma}{8\pi}\int\left[\frac{1+(1-2\cos^2\theta)\beta^2}{ r'^4}\right]{\rm d}V'\nonumber\\
\label{eq:84b.3n}
&=&\frac{Q^{2}\gamma}{2}\left(\frac{1}{a}-\frac{1}{b}\right) \left[1+\frac{\beta^2}{3}\right]\nonumber\\
&=&\frac{Q^2}{2C}\gamma\left[1+\frac{\beta^2}{3}\right]={U}'\gamma\left[1+\frac{\beta^2}{3}\right]\,,
\end{eqnarray} 
\begin{eqnarray}
\label{eq:84b.3m}
{\bf P}_{\rm em}&=&{\frac {1}{c}}\int {T^{10}} {{\rm d}V}\:\hat{\bf x}=\frac{\gamma^2\mbox{\boldmath $\beta$}}{c}\int [T'^{00}+ T'^{11}] \frac{{\rm d}V'}{\gamma}
\nonumber\\
&=&\frac{Q^2}{8\pi}\frac{\gamma\mbox{\boldmath $\beta$}}{c}\int\left[\frac{1+(1-2\cos^2\theta)}{ r'^4}\right]{\rm d}V'\nonumber\\
&=&\frac{Q^{2}}{2}\left(\frac{1}{a} - \frac{1}{b}\right)\frac{\gamma\mbox{\boldmath $\beta$}}{c}\left[1+\frac{1}{3}\right]\nonumber\\
&=&\frac{4}{3}\frac{Q^2}{2C}\frac{\gamma\mbox{\boldmath $\beta$}}{c}=\frac{4}{3}{U}' \frac{\gamma\mbox{\boldmath $\beta$}}{c}\,.
\end{eqnarray} 
Electromagnetic energy and momentum computed thus 
are the same as derived in Eqs.~(\ref{eq:84b.35}) and (\ref{eq:84b.37}). In ${U}_{\rm em}$ (Eq.~(\ref{eq:84b.3n})), the term $\gamma{U}'$ comes from $T^{00}$, while the term $\gamma{U}'{\beta^2}/{3}$ is the contribution of the stress term  $T^{11}$. In the same way, in ${\bf P}_{\rm em}$ (Eq.~(\ref{eq:84b.3m})), the term $\gamma{U}'\mbox{\boldmath $\beta$}/{c}$ comes from $T^{00}$, while the additional term $\gamma{U}'\mbox{\boldmath $\beta$}/{3c}$ arises from the electromagnetic stress term  $T^{11}$. 

It is the contributions of these forces per unit area, represented by the stress terms, $T^{ij}$, in the Maxwell stress tensor (Eq.~(\ref{eq:84b.3g}), that appear in a moving system as additional terms in electromagnetic energy-momentum, in particular, giving rise to the enigmatic factor of 4/3 in the electromagnetic field momentum, even for non-relativistic velocities. 

\section{Contribution of the balancing non-electromagnetic forces -- Poincar\'{e} stresses -- to the energy-momentum}\label{S5}

In a static system, having a net electric charge distribution giving rise to electromagnetic forces within the system, for a {\em static} configuration of the system there would be balancing forces (of non-electromagnetic origin!) in the system that eventually have to be equal and opposite to the forces of electromagnetic interaction everywhere, 
irrespective of the ultimate nature of these balancing forces of constraint. 
Now as seen from another inertial
frame, with respect to which the charged system is in motion,
the work done by the balancing non-electromagnetic forces during a Lorentz contraction 
and the momentum associated with the energy flow due to these non-electromagnetic forces would be equal and opposite to those corresponding to the 
forces of electromagnetic origin, since the two forces are everywhere 
equal and opposite. An inclusion of the contribution of the terms due to the balancing 
non-electromagnetic forces would therefore cancel the corresponding terms 
arising due to the forces of electromagnetic origin (for instance, the 
term on the right hand side of Eqs.~(\ref{eq:84b.54a}) and (\ref{eq:84b.3q})) 
in the momentum and energy of the moving system. As a result, the 4-vector characteristic of 
energy-momentum of the {\em total system}, under a Lorentz transformation, would get restored.

More generally, in a charged system described by an electromagnetic stress-energy tensor, like in the charged spherical capacitor, one can define the Poincar\'{e} stresses \cite{34} by a (mechanical) tensor $M'^{\mu' \nu'}$  
\begin{eqnarray}
\label{eq:84b.3p}
M'^{00} &=& M'^{0i}=M'^{i0}=0\nonumber\,,\\
M'^{ij} &=&-T'^{ij}\,,
\end{eqnarray} 
whose genesis might ultimately be of some otherwise unspecified cause. The implications for the system like a charged spherical capacitor are that in its rest frame all non-electromagnetic forces (Poincar\'{e} stresses!) on the charged plate surfaces are equal and opposite to the electromagnetic forces there, thereby keeping the system in equilibrium. 

The {\em total} stress-energy tensor is the sum of the electromagnetic tensor and the non-electromagnetic tensor, that is $T'^{\mu\nu}+M'^{\mu\nu}$, with only non-zero term in the rest frame being $T'^{00}$, the energy density. Then the total energy density and momentum density, in place of Eqs.~(\ref{eq:84b.3k}) and (\ref{eq:84b.3l}) are instead given by
\begin{eqnarray}
\label{eq:84b.3k1}
\gamma^2\big(T'^{00}+T'^{11}\beta^2+M'^{11}\beta^2\big)=\frac{Q^2\gamma^2}{8\pi r'^4}\,,\\
\label{eq:84b.3l1}
\gamma^2\big(T'^{00} + T'^{11}+ M'^{11}\big)\frac{\beta}{c}=\frac{Q^2\gamma^2}{8\pi r'^4}\frac{\beta}{c}\,,
\end{eqnarray} 
where we have used $T'^{11}+M'^{11}=0$.

A volume integral gives total (electromagnetic + non-electromagnetic) energy-momentum of the charged spherical capacitor in $\cal K$ as
\begin{eqnarray}
\label{eq:84b.3n1}	
{U}_{\rm t}&=&\frac{Q^2\gamma}{8\pi}\int\frac{{\rm d}V'}{ r'^4}=
\frac{Q^{2}\gamma}{2}\left(\frac{1}{a}-\frac{1}{b}\right)\nonumber\\ 
&=& \frac{Q^2\gamma}{2C}={U}'\gamma\,,
\end{eqnarray} 
\begin{eqnarray}
\label{eq:84b.3m1}
{\bf P}_{\rm t}&=&\frac{Q^2}{8\pi}\frac{\gamma\mbox{\boldmath $\beta$}}{c}\int\frac{{\rm d}V'}{ r'^4}=\frac{Q^{2}}{2}\frac{\gamma\mbox{\boldmath $\beta$}}{c}\left(\frac{1}{a}-\frac{1}{b}\right)\nonumber\\ 
&=& \frac{Q^2}{2C}\frac{\gamma\mbox{\boldmath $\beta$}}{c}={U}' \frac{\gamma\mbox{\boldmath $\beta$}}{c}\,.
\end{eqnarray} 
Thus the {\em total} energy and momentum, ${U}_{\rm t}$ and ${\bf P}_{\rm t}$ of the system, like Eqs.~(\ref{eq:84b.38.1}) and Eq.~(\ref{eq:84b.39.1}), form components of a 4-vector under Lorentz transformation.

However, it needs to be emphasized that since the non-electromagnetic interactions, which give rise to Poincar\'{e} stresses, are not represented in electromagnetic fields, and the electromagnetic field energy momentum are still given by Eqs.~(\ref{eq:84b.35}) and (\ref{eq:84b.37}). Therefore the electromagnetic field energy-momentum is not a 4-vector. though the {\em total} energy and momentum, given by Eqs.~(\ref{eq:84b.3n1}) and (\ref{eq:84b.3m1}), do form components of a 4-vector under a Lorentz transformation.
\section{Spherical capacitor electrically uncharged, filled instead with a relativistic perfect fluid}\label{S6}
In a system like a moving perfect fluid, pressure contributes to the momentum of the system, where the fluid could be a liquid, gas or even radiation (photon gas!) \cite{84,MTW73,Sc85}.    
Let $u'$ and $p$ be the uniform energy density and pressure in the rest frame ${\cal K}'$ of the perfect fluid. 

Then in lab frame ${\cal K}$, with respect to which the frame ${\cal K}'$ is moving with a velocity $\bf v$, which could be relativistic, the fluid has an energy density $u$ and a momentum density ${\bf g}$, given by \cite{84,MTW73,La75}
\begin{eqnarray}
\label{eq:84b.4c}
u &=& \gamma^2 \left(u' +p {v^2\over c^2}\right)\:,\\
\label{eq:84b.5c}
{\bf g} &=& \gamma^2\left(u' +p\right) {{\bf v}\over c^2}\: ,
\end{eqnarray}

\begin{figure}
\begin{center}
\includegraphics[width=8cm]{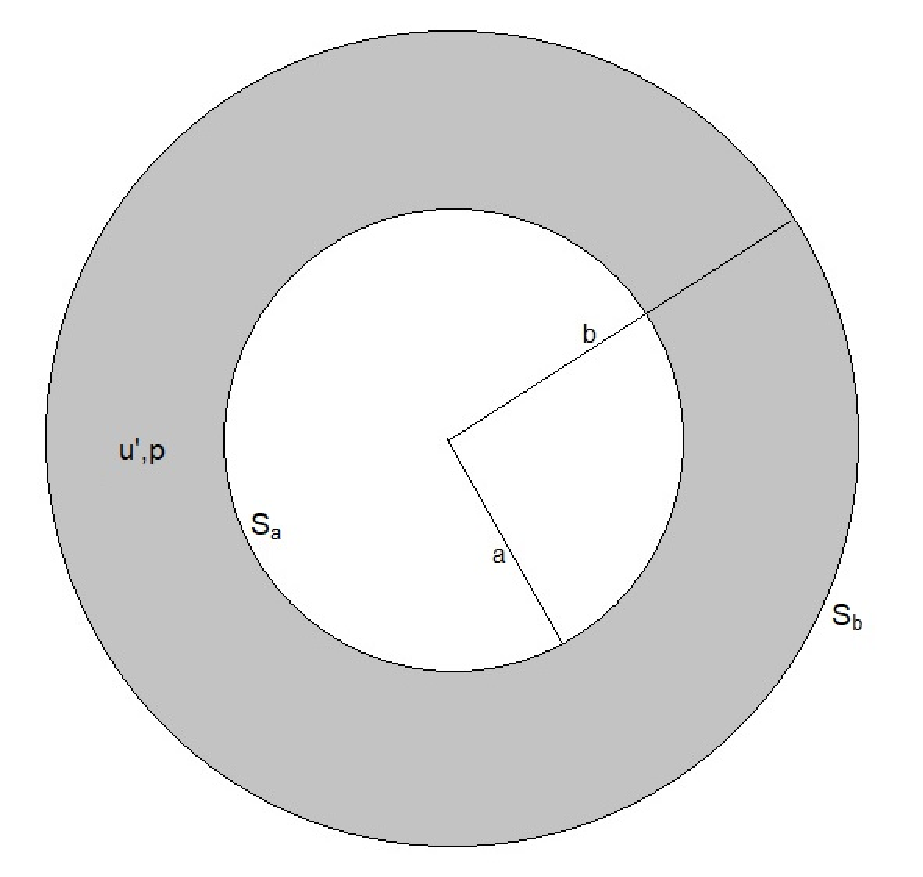}
\caption{A schematic diagram of the spherical capacitor, filled with ultra-relativistic gas molecules or radiation. The capacitor comprises an inner sphere $S_{\rm a}$ of radius $a$ and an outer sphere $S_{\rm b}$ of radius $b$, with the gas or radiation (shown as gray) of energy density $u'$ and pressure $p$ in its rest frame ${\cal K}'$, filling the volume enclosed between the spheres $S_{\rm a}$ and $S_{\rm b}$.}
\label{F3}
\end{center}
\end{figure}

For our purpose, we consider a spherical capacitor, which is electrically uncharged having no electrostatic field inside, but filled instead with a gas having an ultra-relativistic random motion of molecules or an isotropic photon gas a.k.a. radiation,(Fig.~(\ref{F3})). 
Let $u'$ be the uniform energy density of radiation in the rest frame ${\cal K}'$ of the  capacitor, with a corresponding pressure, $p=u'/3$. 
Then from Eqs.~(\ref{eq:84b.4c}) and (\ref{eq:84b.5c}), we have the energy density $u$ and the momentum density ${\bf g}$ in  the lab frame ${\cal K}$, given by \cite{84}
\begin{eqnarray}
\label{eq:84b.4a}
u &=&  \gamma^2 u'\left(1 + {\beta^2\over 3}\right)\:,\\
\label{eq:84b.5a}
{\bf g}&=& \frac{4}{3}\,\gamma^2u'  \:\frac{\mbox{\boldmath $\beta$}}{c} \:.
\end{eqnarray}
The total energy contained in the spherical capacitor system, in rest frame ${\cal K}'$ is 
\begin{eqnarray}
\label{eq:84b.4d}
{U}'&=&u' V'\:, 
\end{eqnarray}
where $V'=4\pi(b^3-a^3)/3$ is the volume enclosed between two spheres $S_{\rm a}$ and $S_{\rm b}$  in the rest-frame.
The momentum of the system in the rest-frame ${\cal K}'$ is zero, of course. 
\begin{eqnarray}
\label{eq:84b.5d}
{P}'&=&0\:. 
\end{eqnarray}

Seen from the lab frame ${\cal K}$, the energy and momentum of the system are \cite{84}
\begin{eqnarray}
\label{eq:84b.4f}
{U}&=& \gamma {U}' \left(1 +{\beta^2\over 3}\right)\:,\\
\label{eq:84b.5f}
{\bf P}&=& \frac{4}{3} \gamma {U}'\:\frac{\mbox{\boldmath $\beta$}}{c}\:. 
\end{eqnarray}
While for an individual photon, energy and momentum do transform as components of a 4-vector, for an isotropic velocity distribution in a photon gas, energy and momentum of the total  gas transforms as in Eqs.~(\ref{eq:84b.4f}) and (\ref{eq:84b.5f}). 
In fact, these expressions can be obtained directly by first computing in lab frame ${\cal K}$ energy and momentum of individual photons moving in random direction in  frame ${\cal K}'$, using the aberration and Doppler shift formulas and then integrating over the transformed number distribution one arrives at the Eqs.~(\ref{eq:84b.4f}) and (\ref{eq:84b.5f}) \cite{85}.

The transformation relations for the energy and momentum of the radiation within the system thus 
are exactly the same as derived in Eqs.~(\ref{eq:84b.35}) and (\ref{eq:84b.37}) or in Eqs.~(\ref{eq:84b.3n}) and (\ref{eq:84b.3m}).

Now for a non-relativistic motion ($v\ll c$ and $\gamma \approx 1$), we have ${U}= {U}'$, while for momentum we get
\begin{eqnarray}
\label{eq:84b.5g}
{\bf P}&=& \frac{4}{3}\,{U}'\: \frac{\mbox{\boldmath $\beta$}}{c} \:,
\end{eqnarray}
It should be noted here that it is the {\em bulk} velocity $\bf v$ of the fluid (or rather of its container like the spherical capacitor), that could be non-relativistic, however, molecules of the gas may be randomly moving with ultra-relativistic speeds or the photons moving in random directions with $c$, In relativistic fluids, it is $(u' + p)/c^2$, in place of $u'/c^2$, that plays the role of `inertial mass density', in that, the larger $(u' + p)$ is, the harder it becomes to accelerate the object \cite{Sc85}. In highly relativistic fluids, the mass density becomes $4u'/3c^2$.

Thus one arrives at a factor of 4/3 in the momentum of a spherical capacitor filled with an ultra-relativistic gas or  radiation, and having no electric charges or any consequential electrostatic fields in the rest frame. Thus the ``enigmatic''  factor of 4/3 arises from the contribution of forces or stress whose genesis in the system may be of non-electromagnetic origin, and has less to do with the presence of electric charges or currents within the system. No modifications in the definitions of energy-momentum of moving fluids have ever been found necessary to be proposed; physics after all should be the same in the case of electromagnetic energy-momentum as well.  Also if we let $a \rightarrow 0$, then our radiation-filled capacitor becomes a radiation-filled sphere, with the same momentum relation (Eq.~(\ref{eq:84b.5g})) as a charged sphere \cite{84,15}.
Thus, in a spherical system, like a spherical shell or a spherical capacitor comprising two concentric spherical shells, where the volume enclosed is filled with ultra-relativistic gas or a photon gas, exactly the same 4/3 factor in momentum appears (Eq.~(\ref{eq:84b.5g})) as in the case of an electrically charged sphere on a microscopic scale (classical electron model) or a charged spherical capacitor on a macroscopic scale. In all cases the apparently anomalous factor of 4/3 in momentum for a moving system, arises from the contribution of pressure or stress, whatever might be the source of that stress in the system, whether electromagnetic or non-electromagnetic. Of course, in non-spherical systems, there might be some other extra factor instead of 4/3, e.g., in a charged parallel plate capacitor, the extra factor is two for a motion parallel to the plate surfaces while for a motion perpendicular to the plate surfaces, the factor is zero with the momentum becoming null for a moving charged capacitor or a similar capacitor filled with a non-electric fluid \cite{84,15}.

It has been said that this 4/3 factor in momentum appears only in the so-called  ``bound'' fields, associated with electric charges, and not in ``free'' fields representing electromagnetic radiation \cite{2,6,Er02}  and the modified definitions of electromagnetic energy-momentum, accordingly, have been proposed for the bound fields alone, to eliminate this unwanted factor of 4/3. The conventional wisdom thus is that the energy-momentum of free electromagnetic radiation, contained in a volume having no electric charges or currents, transforms as a 4-vector, without any such 4/3 factor \cite{2}.  On the other hand, we have demonstrated here, that the factor of 4/3 in momentum appears as much for radiation fields where pressure plays a significant role in the momentum formulation, and that this factor is not something arising exclusively from electric charges in the system and there is nothing amiss in the standard definition of electromagnetic energy-momentum.

Of course, if in the lab frame ${\cal K}$, we as well include in the energy and momentum, the contributions of the mechanical forces, akin to Poincar\'{e} stresses, of surfaces  $S_{\rm a}$ and $S_{\rm b}$, that keep the radiation confined within the volume between $S_{\rm a}$ and $S_{\rm b}$, then we do get {\em total} energy and momentum which transform as components of a 
4-vector.

\section*{Data Availability}
Data sharing not applicable to this article as no datasets were generated or analysed during the current study.
\section*{Declarations}
The author has no conflicts of interest/competing interests to declare that are relevant to the content of this article. No funds, grants, or other support of any kind was received from anywhere for this research.
{}
\end{document}